\documentclass[prb,twocolumn,showpacs,preprintnumbers]{revtex4}

\usepackage{graphicx}

\begin{document}


\title{Time evolution and decoherence of entangled states realized in 
coupled superconducting flux qubits}

\author{Takuya Mouri\({}^{1,2}\),\hspace{3mm}Hayato 
Nakano\({}^{2,3}\),\hspace{3mm}and Hideaki Takayanagi\({}^{1,2,3}\)}
 
\affiliation{\({}^1\)Department of Physics, Tokyo University of 
Science, Tokyo 162-8601, Japan\\ \({}^2\)CREST, Japan Science 
Technology Agency, Saitama 332-0012, Japan\\ \({}^3\)NTT Basic 
Research Laboratories, NTT Corporation, Kanagawa 243-0198, Japan }

\date{\today}

\begin{abstract}
We study theoretically how decoherence affects superposition states 
composed of entangled states in inductively coupled two 
superconducting flux-qubits. We discover that the quantum fluctuation 
of an observable in a coupled flux-qubit system plays a crucial role 
in decoherence when the expectation value of the observable is zero. 
This examplifies that decoherence can be also induced through a 
quantum mechanically higher-order effect. We also find that there 
exists a decoherence free subspace for the environment coupled via a 
charge degree of freedom of the qubit system. 
\end{abstract}

\maketitle

\section{Introduction}
Josephson-junction circuits behave quantum mechanically, if they are 
sufficiently decoupled from their environment. However, in reality, 
since a macroscopic quantum state is difficult to be decoupled from 
their environment completely, a quantum mechanical superposition 
state suffers decoherence. This is one of the central problems that 
must be solved before these circuits can be used for quantum 
information processing[1]. In order to minimize decoherence on 
quantum states, it is very important to search what is the dominant 
source of decoherence and how sensitively qubits feel the effect of 
environment. In this paper, we theoretically present that the quantum 
fluctuation of an observable in a coupled flux-qubit plays a crucial 
role in decoherence, even if the expectation value of the obseravable 
is always zero. We also report that we can extract a decoherence-free 
single-qubit basis from an inductively coupled flux-qubit system. 

\section{Inductively coupled two flux qubit system}
Among qubits based on Josephson-junction circuits, a flux qubit was 
realized as a superconducting ring with three 
Josephson-junctions[2,3] and coherent oscillations were 
demonstrated[4,5,6]. We discuss decoherence appearing in the dynamics 
of an inductively coupled  two superconducting flux-qubits in Fig. 
1(a)[7,8]. Four of Josephson-junctions have the Josephson energy 
\(E_{J}\) and capacitance \(C\). The remaining two have 
\({\alpha}E_{J}\) and \({\alpha}C\), where \(\alpha=0.75\) is a 
constant. The Hamiltonian of the system is given by 
\begin{eqnarray}
\displaystyle{
H=\frac{C}{2}\left(\frac{\hbar}{2e}\right)^2\left({\dot\gamma_{L1}}^2+{\dot\gamma_{R1}}^2+{\dot\gamma_{L2}}^2
+{\dot\gamma_{R2}}^2+\alpha({\dot\gamma_{L3}}^2+{\dot\gamma_{R3}}^2)\right)} \nonumber\\
-E_{J}(\cos[\gamma_{L1}]+\cos[\gamma_{R1}]+\cos[\gamma_{L2}]
+\cos[\gamma_{R2}]\\+\alpha(\cos[\gamma_{L3}]+\cos[\gamma_{R3}]))
+\frac{1}{2M}\left(\frac{\hbar}{2e}\right)^2\gamma_{M}^2,\nonumber
\end{eqnarray}
where \(\gamma_i\) \((i=L1,R1,L2,R2,L3,R3,M)\) is the phase 
difference across each junction or the inductance. By calculating the 
Hamiltonian with parameters \(E_{J}=100{\rm GHz}\), 
\(E_{C}=\frac{(2e)^2}{2C}=8{\rm GHz}\), \(\alpha=0.75\), 
\(M=\frac{1}{20E_{J}}(\frac{\hbar}{2e})^2\), we can obtain the energy 
dispersion shown in Fig. 1(b) and we call the point at \(f=0.5\) as 
the ``degeneracy point".

\begin{figure}[ht]
\includegraphics[width=8.5cm]{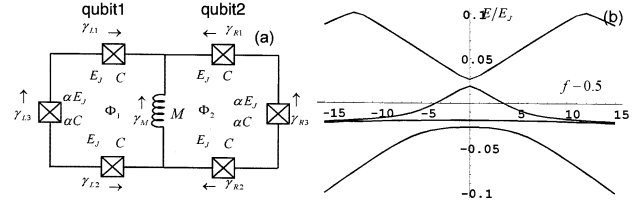}
\caption{\label{fig:epsart1}(a)Effective circuit of inductively 
coupled two flux qubits. Two superconducting closed loops correspond 
to two flux qubits. A qubit has three Josephson junctions. An extenal 
flux \(\Phi_{1}\) or \(\Phi_{2}\) is penetrating through each qubit 
loop. Qubits are coupled to each other through the mutual inductance 
\(M\). (b)Energy dispersions obtained when identical external fluxes 
are applied to the qubits, \(f={\Phi_1}/{\Phi_0}={\Phi_2}/{\Phi_0}\) 
(\(\Phi_0={h}/{2e}\)).}
\end{figure}

We can approximate the Hamiltonian for the two-qubit system using 
Pauli matrices \(\sigma_i^z\), \(\sigma_i^x\),
\[H_{\rm 
S}=h_1\sigma_1^z+\Delta_1\sigma_1^x+h_2\sigma_2^z+\Delta_2\sigma_2^x+j\sigma_1^z\sigma_2^z.\]

Here, subscripts 1 and 2 indicate the left and right qubits, 
respectively. The basis of the two-qubit state are 
\(|\downarrow_1\downarrow_2\rangle\),\(|\downarrow_1\uparrow_2\rangle\),\(|\uparrow_1\downarrow_2\rangle\) 
and \(|\uparrow_1\uparrow_2\rangle\), where \(|\uparrow_i\rangle\) 
corresponds to the state where the persistent current flows 
counterclockwise and \(|\downarrow_i\rangle\) to the clockwise state 
around the loop of the qubit \(i\)(\(i=1,2)\). \(h_i\)(\(i=1,2)\) is 
the parameter that depends on the external flux \(\Phi_i\) for the 
qubit \(i\), and 
\(h_i=I_{pi}(\Phi_i-\frac{\Phi_0}{2})=I_{pi}\Phi_0(f_i-\frac{1}{2})\), 
where \(I_{pi}\)(\(i=1,2\)) is the absolute value of the persistent 
current in the each qubit determined by \(E_J\) and \(\alpha\). 
\(\Phi_0={h}/{2e}\) is the flux quantum. \(\Delta_i\)(\(i=1,2\)) is 
the tunneling matrix element each qubit has, which is the order of 
\(\sqrt{E_CE_J}e^{-\sqrt{E_J/E_C}}\). \(j\) is the coupling constant 
between two qubits, which is determined approximately by 
\(MI_{p1}I_{p2}\). If the external flux is equally applied 
\((h_1=h_2=h)\) and the two qubits are assumed to be identical 
\((\Delta_1=\Delta_2=\Delta)\), we can simplify the Hamiltonian as 
\[H_{\rm 
S}=h\sigma_1^z+\Delta\sigma_1^x+h\sigma_2^z+\Delta\sigma_2^z+j\sigma_1^z\sigma_2^z.\]

By comparing energy eigenstates of the \(H_{\rm S}\) and Fig. 1(b) 
near the degeneracy point, we find that this approximated Hamiltonian 
\(H_{\rm S}\) reproduces well the result of the original two 
flux-qubit Hamiltonian with \(j=1.78{\rm GHz}\) and \(\Delta=2.07{\rm 
GHz}\). Therefore, hereafter we analyze the decoherence in our 
coupled flux-qubit system with this \(H_{\rm S}\). In terms of the 
original flux-qubit system, \(\sigma_i^z\) corresponds to the flux 
the qubit \(i\) has, and  \(\sigma_i^x\) to the charge polarization 
in the junction of the qubit. At the degeneracy point \(h=0\), we can 
obtain eigenvalues and eigenstates[7],
\begin{eqnarray}
E_0=-\sqrt{j^2+4\Delta^2},\hspace{5mm}|\psi_0\rangle=|\uparrow_1\uparrow_2\rangle+|\downarrow_1\downarrow_2\rangle 
\nonumber\\
-(\frac{j+\sqrt{j^2+4\Delta^2}}{2\Delta})(|\uparrow_1\downarrow_2\rangle+|\downarrow_1\uparrow_2\rangle), 
\nonumber\\
E_1=-j,\hspace{5mm}|\psi_1\rangle=|\uparrow_1\downarrow_2\rangle-|\downarrow_1\uparrow_2\rangle, 
\nonumber\\
E_2=j,\hspace{5mm}|\psi_2\rangle=|\uparrow_1\uparrow_2\rangle-|\downarrow_1\downarrow_2\rangle, 
\nonumber\\
E_3=\sqrt{j^2+4\Delta^2},\hspace{5mm}|\psi_3\rangle=|\uparrow_1\uparrow_2\rangle+|\downarrow_1\downarrow_2\rangle 
\\
-(\frac{j-\sqrt{j^2+4\Delta^2}}{2\Delta})(|\uparrow_1\downarrow_2\rangle+|\downarrow_1\uparrow_2\rangle).
\nonumber
\end{eqnarray}
Here, a remarkable fact is that all the eigenstates of the two-qubit 
Hamiltonian at the degeneracy point are strongly entangled ones. 
Adopting a state composed of these entangled states as an initial 
state, we will discuss the time evolution of the density operator and 
some quantities.

\section{physical model of decoherence and Quantum master equation}
In order to take into account decoherence, we consider the total 
system as the sum of the qubit system and the environment. The total 
Hamiltonian is \(H_{\rm TOT}=H_{\rm S}+H_{\rm E}+H_{\rm INT}\), where 
\(H_{\rm S}\), \(H_{\rm E}\), and \(H_{\rm INT}\), are the 
Hamiltonian of the qubit system, the environment, and the interaction 
between them, respectively. We showed already \(H_{\rm S}\) and the 
components of environment and interaction are described as follows; 
\(H_{\rm E}=\sum_{n} 
\hbar\omega_na_n^{\dagger}a_n\),\hspace{3mm}\(H_{\rm 
INT}=x\sum_k\lambda_{k}(a_k+a_k^{\dagger})\). We regard the 
environment as a boson bath. \(a_k\), \(a_k^{\dagger}\) are the 
annihilation and the creation operators of the environmental mode 
\(k\). \(\lambda_k\) is the strength of the coupling in each mode. 
\(x\) means a physical quantity of the qubit system that is coupled 
to the environment. Assuming that the system-environment coupling is 
small enough, we can treat \(H_{\rm INT}\) as a perturbative term so 
that we acquire a quantum master equation[9],
\begin{eqnarray}
\dot \rho_{\rm S}=\frac{1}{\hbar}\int_0^t{\rm 
d}t_1(\nu(t_1)[x,[x(-t_1),\rho_{\rm S}]] \nonumber\\
-i\eta(t_1)[x,\{x(-t_1),\rho_{\rm S}\}]), \nonumber
\end{eqnarray}
where
\begin{eqnarray}
\nu(t)= \int_0^{\infty}{\rm 
d}\omega\cos[{\omega}t]\coth[\frac{\omega}{2k_BT}]J(\omega), \nonumber\\
\eta(t) =\int_0^{\infty}{\rm d}\omega\sin[{\omega}t]J(\omega).\nonumber
\end{eqnarray}
\(k_B\) is the Boltzman constant and \(T\) is the absolute 
temperature. We adopt the spectral density 
\(J(\omega)=g\frac{{\omega}a^2}{a^2+{\omega^2}}\). Here \(a\) is the 
cut off frequency and we set \(a=5\).

\section{Results and discussion}
\subsection{decoherence via flux degree}
First, we assume that the interaction between the system and the 
environment is caused only through the total flux of qubits; then we 
take \(x=\sigma_1^z+\sigma_2^z\). This is a reasonable situation 
because we can expect that the system suffers decoherence due to the 
coupling to the detector like a SQUID and the fluctuation of the 
external flux, both of which are related to small flux generated by 
the persistent currents flowing in qubits. Now, we consider a 
superposition state composed of the first and second excited states 
\(|\psi_{12}(0)\rangle=\frac{1}{\sqrt{2}}(|\psi_1\rangle+|\psi_2\rangle)\) 
at the degeneracy point\((h=0)\) as the initial state. We can 
analytically calculate the time evolutions of expectation values 
\(\langle\sigma_1^z\rangle\), \(\langle\sigma_2^z\rangle\), 
\(\langle\sigma_1^x\rangle\) and \(\langle\sigma_2^x\rangle\) when 
there is no decoherence. Provided that it is possible to set 
independent probes which can detect both flux and charge on each of 
qubits, we can measure expectation values 
\(\langle\sigma_1^z\rangle\), \(\langle\sigma_2^z\rangle\), 
\(\langle\sigma_1^x\rangle\) and \(\langle\sigma_2^x\rangle\) 
independently. \(|\psi_{12}(t)\rangle\) and the time evolutions of 
observables at \(h=0\) are obtained as follows;

\begin{eqnarray}
|\psi_{12}(t)\rangle=\cos[jt]|\downarrow_{x1}\uparrow_{x2}\rangle+i{\rm 
sin}[jt]|\uparrow_{x1}\downarrow_{x2}\rangle, \\
\langle\psi_{12}(t)|\sigma_1^z|\psi_{12}(t)\rangle=0, \nonumber\\
\langle\psi_{12}(t)|\sigma_2^z|\psi_{12}(t)\rangle=0, \nonumber\\
\langle\psi_{12}(t)|\sigma_1^x|\psi_{12}(t)\rangle=-\cos[2jt], \nonumber\\ 
\langle\psi_{12}(t)|\sigma_2^x|\psi_{12}(t)\rangle=\cos[2jt].\nonumber
\end{eqnarray}
Here 
\(|\uparrow_{xi}\rangle=|\uparrow_i\rangle+|\downarrow_i\rangle\),\hspace{3mm}\(|\downarrow_{xi}\rangle=-|\uparrow_i\rangle+|\downarrow_i\rangle\)\hspace{3mm}(\(i=1,2)\). 
This shows that this coherent oscillation is impossible to detect 
through the flux, but there is a possiblity to detect through the 
charge. We will show the result of the numerical calculation in the 
presence of decoherence, which is obtained by solving the quantum 
master equation with the above initial state. As we have expected in 
the analytical calculation, we realize that expectation values 
\(\langle\sigma_1^z\rangle\) and \(\langle\sigma_2^z\rangle\) 
obtained in numerical calculations are always zero. The most 
interesting feature of this example is that the expectation value of 
the physical quantity which interacts with the environment 
\(\langle\sigma_1^z+\sigma_2^z\rangle\) is zero. Although we would 
speculate that there is no coupling between qubits and the 
environment, this turns out not true according to the calculation 
results. 
\begin{figure}[ht]
\includegraphics[width=8.5cm]{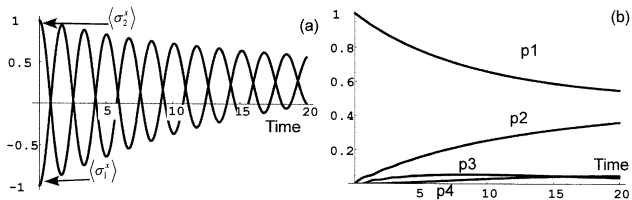}

\caption{\label{fig:epsart2}(a)Time evolution of expectation values 
\(\langle\sigma_1^x\rangle\) and \(\langle\sigma_2^x\rangle\) in the 
presence of decoherence. (b)Time evolution of the eigenvalues 
\(p_i\)\((i=1,2,3,4)\) of the density operator. (\(h=0\), \(j=1.78\), 
\(\Delta=2.07\), \(g=1\)) Time is normalized by \(\Delta\). The 
initial state is \(|\psi_{12}(0)\rangle\). Decoherence is caused 
through \(x=\sigma_1^z+\sigma_2^z\).}
\end{figure} 
Figure 2(a) shows the oscillations of \(\langle\sigma_1^x\rangle\) 
and \(\langle\sigma_2^x\rangle\). We find that thier amplitudes are 
obviously suffering decoherence. Time evolution of eigenvalues of the 
density operator in Fig. 2(b)  helps us realize that a pure state is 
rapidly changing into a mixed state due to decoherence because the 
emergence of the finite second and later eigenvalues means the system 
becomes a mixed state. Therefore, we can make sure that, even if the 
expectation value of the interacting quantity between system and 
environment is always zero, there exists certain effect of 
decoherence.

Next, we discuss when the initial state is the superposition state 
\(|\psi_{01}(0)\rangle=\frac{1}{\sqrt{2}}(|\psi_0\rangle+|\psi_1\rangle)\) 
of the ground and the first excited states. 
\begin{figure}[ht]
\includegraphics[width=8.5cm]{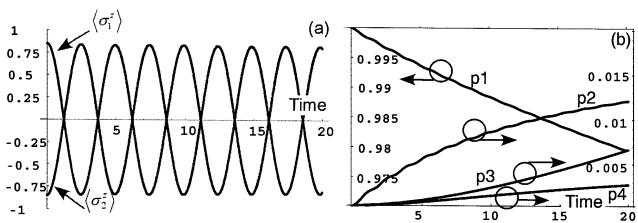}
\caption{\label{fig:epsart3}(a)Time evolution of expectation values 
\(\langle\sigma_1^z\rangle\) and \(\langle\sigma_2^z\rangle\) with 
decoherence via \(x=\sigma_1^z+\sigma_2^z\). (b)Time evolution of 
eigenvalues of the density operator. The initial state is 
\(|\psi_{01}(0)\rangle\).}
\end{figure}
Figure 3(a) shows oscillations of the flux induced at the qubit1 and 
qubit2, and these oscillations suggest that the possibility to detect 
the fluxes \(\langle\sigma_1^z\rangle\), \(\langle\sigma_2^z\rangle\) 
separately by setting an independent SQUIDs over each of the qubits. 
Figure 3(a) also shows that amplitudes of oscillations hardly decay 
and it seems that the oscillations are hardly affected by 
decoherence. This suggests that it could be possible to realize 
coherent oscillation which is robust against the decoherence via 
\(x=\sigma_1^z+\sigma_2^z\). We can again observe the expectation 
value of the coupling between the qubit system and the environment is 
always zero (\(\langle\sigma_1^z+\sigma_2^z\rangle=0)\). From the 
above two examples, surely we can make coherent oscillations where 
\(\langle{H_{\rm 
INT}}\rangle=\langle{x}\rangle\sum_k\lambda_{k}(a_k+a_k^{\dagger})=0\) 
by using initial states which is composed by entangled energy 
eigenstates. However, one is strongly affected by decoherence and the 
other is not. Then, we are interested in considering how quantum 
fluctuations of the coupling \(x\) may become the source of 
decoherence[8]. In fact, the calculations obtaining Figs. 2 and 3 
also give the finite quantum fluctuations 
\begin{eqnarray}
\sqrt{\langle\psi_{12}(t)|\delta{x^2}|\psi_{12}(t)\rangle}\sim1.4, 
\nonumber\\
\sqrt{\langle\psi_{01}(t)|\delta{x^2}|\psi_{01}(t)\rangle}\sim0.75,\nonumber
\end{eqnarray}
where \(\delta{x^2}{\equiv}x^2-{\langle{x}\rangle}^2\). It looks like 
that the quantum fluctuations of the coupling \(x\) and resulting 
magnitude of decoherence are determined by the curvatures of energy 
dispersions in Fig. 1(b) around the degeneracy point \(f=0.5\). The 
quantum fluctuation of \(x=\sigma_1^z+\sigma_2^z\) qualitatively 
corresponds to the quantum fluctuation of an applied flux in the 
horizontal axis of Fig. 1(b). In fact, by a perturbation calculation 
under the condition \(|h|\ll|\Delta|,|j|\), the energy dispersion of 
an eigenstate \(|i\rangle\) in the vicinity of the degeneracy point 
is given by 
\[E_{i}{\sim}E_{i}^{0}-{h^2}\sum_{{j}\neq{i}}\frac{{\langle{i0}|\sigma_1^z+\sigma_2^z|{j0}\rangle}{\langle{j0}|\sigma_1^z+\sigma_2^z|{i0}\rangle}}{E_j^0-E_i^0},\]

where \(E_i^0\) and \(|i0\rangle\) are the eigenenergy and eigenstate 
at the degeneracy point. So, the curvature 
\(\frac{\partial^2E_i}{\partial{h}^2}\)(\(i=0,1,2\)) is approximately 
proportional to 
\[\sum_{j}{\langle{i0}|\sigma_1^z+\sigma_2^z|{j0}\rangle}{\langle{j0}|\sigma_1^z+\sigma_2^z|{i0}\rangle},\]

that is the quantum fluctuation of \(\sigma_1^z+\sigma_2^z\). As a 
result, we have reached a conclusion that the decoherence at the 
degeneracy point depends on the curvature of the energy dispersion 
around the degeneracy point. 

Now we discuss the concurrence which is defined as 
\(C(t)=\sqrt{\xi_1}-\sqrt{\xi_2}-\sqrt{\xi_3}-\sqrt{\xi_4}\) where 
\(\xi_i\)\((i=1,2,3,4)\) is the \(i\) th eigenvalue of 
\(\Lambda(t)\equiv\sum_i\left|p_i\right|^2\left|\langle\psi_i|\sigma_1^y\sigma_2^y|\psi_i\rangle\right|^2\). 
This is a very important measure because the concurrence tells us how 
the two qubits keep entanglement which provide us the possiblity to 
implement quantum parallelism calculations. Figure 4 shows the time 
evolution of the concurrence. When compared with the decays of the 
corresponding coherent oscillations in Figs. 2(a) and 4(a), we can 
find that the entanglement decays more rapidly than the observables 
\(\langle\sigma_1^x\rangle\) and \(\langle\sigma_2^x\rangle\)in the 
presence of decoherence.

\begin{figure}[ht]
\includegraphics[width=8.5cm]{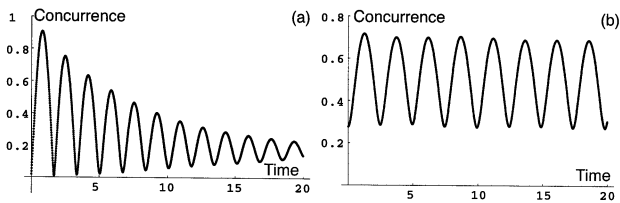}
\caption{\label{fig:epsart4}(a)Time evolution of the concurrence with 
the initial state \(|\psi_{12}(0)\rangle\). (b)Time evolution of the 
concurrence with the initial state \(|\psi_{01}(0)\rangle\).}
\end{figure}

\subsection{decoherence via charge degree}

Next, we suppose that the interaction between the system and the 
environment is only through the total charge freedom the qubits have; 
\(x=\sigma_1^x+\sigma_2^x\). For example, this type of interaction 
might appear in the presence of residual charges near the qubit 
junctions. When the initial state is 
\(|\psi_{12}(0)\rangle=\frac{1}{\sqrt{2}}(|\psi_1\rangle+|\psi_2\rangle)\), 
as we have analytically calculated, we have already known 
\(\langle\sigma_1^z\rangle\) and \(\langle\sigma_2^z\rangle\) are 
always zero and 
\(\langle\psi_{12}(t)|\sigma_1^x|\psi_{12}(t)\rangle\) and 
\(\langle\psi_{12}(t)|\sigma_2^x|\psi_{12}(t)\rangle\) oscillate 
sinusoidally. The numerical calculation of the time evolution gives 
the behavior of the expectation values, shown in Fig. 5(a). We can 
see \(\langle\sigma_1^x\rangle\) and \(\langle\sigma_2^x\rangle\) 
amazingly keep oscillating without any reduction of the amplitudes, 
and we have confirmed that the pure state is kept perfectly. We find 
this is because the two basis states \(|\psi_1\rangle\) and 
\(|\psi_2\rangle\) composing the initial state 
\(|\psi_{12}(0)\rangle\) construct a Decoherence Free 
Subspace(DFS)[10] for the coupling between the two qubit system and 
the environment via \(x=\sigma_1^x+\sigma_2^x\)[10,11]. In other 
words, \(|\psi_1\rangle\) and \(|\psi_2\rangle\) are the degenerate 
eigenstates of \(x\). Therefore, we can suggest that it is possible 
to design a {\it single} quantum bit which does not form entanglement 
with the environment through the charge degree of freedom for a 
superposition state of \(|\psi_1\rangle\) and \(|\psi_2\rangle\). 
This is an example of a suppression of decoherence by introducing a 
redundant qubit and a qubit-qubit interaction, like quantum error 
correction codes[12]. Figure 5(b) represents time evolution of the 
concurrence and we can make sure that the entanglement never 
deteriorates because of the suppresion of decoherence in the DFS.

\begin{figure}[h]
\includegraphics[width=8.5cm]{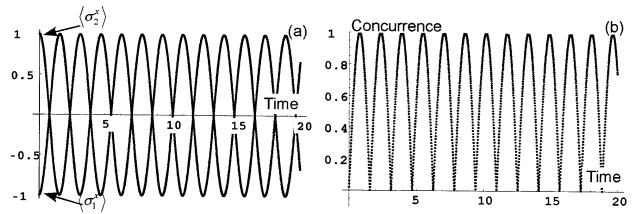}
\caption{\label{fig:epsart5}(a)Time evolution of the expectation 
values \(\langle\sigma_1^x\rangle\), \(\langle\sigma_2^x\rangle\). 
(b)Time evolution of concurrence. The initial state is 
\(|\psi_{12}(0)\rangle\). Decoherence is caused through 
\(x=\sigma_1^x+\sigma_2^x\).}
\end{figure}

\section{summary and conclusion}
In summary, we analyzed the time evolution of an inductively coupled 
flux qubit system in the presence of environment. We confirmed the 
importance of quantum fluctuations of the observable coupled to the 
environment. We also found two basis states constructing a DFS for 
charge fluctuations.

\section*{Acknowledgements}
We thank F. Wilhelm for valuable discussions. We also acknowledge K. 
Semba, S. Saito, H. Tanaka, F. Deppe and J. Johansson for their 
comments from the viewpoint of experiments. M. Ueda should be 
acknowledged for his helpful advices.


\begin{thebibliography}{99}
\bibitem{B1}Y. Makhlin, G. Sch\"{o}n and A. Shnirman, Rev. Mod. Phys. 
357, 73, (2001), and references therein.
\bibitem{B2}J. E. Mooij, T.P. Orlando, L. Levitov, Lin. Tian, Casper 
H. van der Wal, Seth Lloyd, Science {\bf 285}, 1036 (2003).
\bibitem{B3}C. H. van der Wal, A. C. J. ter Haar, F. K. Wilhelm, R. 
N. Schouten, C. J. P. M. Harmans, T. P. Orlando, Seth. Lloyd, J. E. 
Mooij, Science {\bf 290}, 773 (2000).
\bibitem{B4}I. Chiorescu, Y. Nakamura, C. J. P. M. Harmans, and J. E. 
Mooij, Science \(\bf 299\), 1869 (2003).
\bibitem{B5}H. Tanaka {\it et al.\/}, cond-mat/0407299.
\bibitem{B6}T. Kutsuzawa {\it et al.\/}, submitted to Appl. Phys. 
Lett.
\bibitem{B7}J. B. Majer, F. G. Paauw, A.C. ter Haar, C.J.P.M. 
Harmans, and J. E. Mooij, cond-mat/0308192.
\bibitem{B8}J. Q. You, Y. Nakamura, and Franco. Nori, 
cond-mat/0309491.
\bibitem{B9}J.P. Paz and W.H. Zurek, p. 536 ineCoherent atomic 
matter waves', edited by P. Kaiser {\it et al.\/}, (Springer-Verlag, 
Berlin, 2001).
\bibitem{B10}D. A. Lidar and K. B. Whaley, p. 83 ineIrreversible 
Quantum Dynamics', edited by F. Benatti, R. Floreanini, 
(Springer-Verlag, Berin, 2003).
\bibitem{B11}A similar DFS was discovered in a charge-coupled flux 
qubits. M.J. Storcz {\it et al.\/}, cond-mat/0407780.
\bibitem{B12}H. Nakano and H. Takayanagi, p. 28 in Proc. of 
ISQM-Tokyo'01, edited by Y. A. Ono and K. Fujikawa (World Scientific, 
Singapore, 2002).



\end{thebibliography}
\end{document}